\documentclass[doublespacing]{elsart}

\usepackage{amssymb}
\usepackage{amsmath}
\usepackage{bm}
\usepackage{graphicx}

\begin{document}

\begin{frontmatter}



\title{Quadrupole Moment of $^{37}$K}


\author[nscl]{K. Minamisono}\ead{minamiso@nscl.msu.edu}, 
\author[nscl,chem]{P. F. Mantica},
\author[nscl,chem]{H. L. Crawford}, 
\author[nscl,chem]{J. S. Pinter},
\author[nscl,chem]{J. B. Stoker}, 
\author[nscl,jaeri]{Y. Utsuno}, and 
\author[nscl,chem]{R. R. Weerasiri\thanksref{label1}}

\address[nscl]{National Superconducting Cyclotron Laboratory, Michigan State University, East Lansing, MI 48824, USA}
\address[chem]{Department of Chemistry, Michigan State University, East Lansing, MI 48824, USA}
\address[jaeri]{Japan Atomic Energy Agency, Tokai, Ibaraki 319-1195, Japan}

\thanks[label1]{Current address: Technova Corp. Lansing, MI 48906, USA}

\begin{abstract}
The electric quadrupole coupling constant of the ground state of $^{37}$K($I^{\pi}$ = 3/2$^+$, $T_{1/2}$ = 1.22 s) in a tetragonal KH$_2$PO$_4$ single crystal was measured to be $|eqQ/h| = 2.99 \pm 0.07$ MHz using the $\beta$-ray detecting nuclear quadrupole resonance technique.  The electric quadrupole moment of $^{37}$K was determined to be $|Q(^{37}{\rm K})| = 10.6 \pm 0.4$ efm$^2$, where the known electric quadrupole coupling constant of stable $^{39}$K in the KH$_2$PO$_4$ crystal was used as a reference.  The present experimental result is larger than that predicted by shell-model calculations in the $sd$ or the $sd$ and $fp$ model spaces.  A possible variation of effective charges was explored to explain the discrepancy.
\end{abstract}

\end{frontmatter}

\section{Introduction}
\label{introduction}
The spectroscopic electric-quadrupole moment ($Q$) provides a direct measure of the deviation of the charge distribution in a nucleus from spherical symmetry.  Proton and neutron effective charges are used in shell-model calculations to obtain quadrupole moments.  Within a given shell-model space, the effective charges represent 2$\hbar \omega$, $I^{\pi}$ = 2$^+$ excitations of the core nucleons to valence orbits \cite{sagawa84} and reflect the virtual excitation of the isoscalar and isovector giant quadrupole resonances.  Values of effective charges for $sd$-shell nuclei have been obtained from a systematic analysis of experimental $E$2 matrix elements \cite{sagawa84, brown88} and give reasonable agreement between experimental and theoretical quadrupole moments.  However, some nuclei show significant disagreement between experiment and theory.  The neutron-rich B isotopes, for example, show a reduction of the neutron effective charge \cite{ogawa04} because the loosely-bound valence neutrons, far removed from the core, have less probability to excite the core than well bounded neutrons.  Neutron-deficient nuclei, especially those with small proton separation energies, may also be expected to show variation of effective charges.  Experimental $Q$ of such neutron-deficient nuclei are still scarce, even in the $sd$ shell.  Additional systematic data of $Q$ of neutron-deficient nuclei are important to further improve our knowledge about the exotic structure of dripline nuclei.  



The previously measured quadrupole moment of neutron-deficient $^{37}$K, $Q$($^{37}$K) = 11 $\pm$ 4 efm$^2$, was obtained from the hyperfine coupling constant of the $4p^2$ $P_{3/2}$ state of the $^{37}$K atom \cite{behr97}.  In the evaluation of $Q(^{37}$K),  the $4p^2$ $P_{3/2}$ coupling constant of the $^{39}$K \cite{arimondo77} atom and $Q(^{39}$K) evaluated with a calculated electric field gradient \cite{sundholm93} were used as reference.  The OXBASH shell-model calculation code \cite{brown} in the $sd$ model space with the USDA interaction \cite{brown06} gives $Q$($^{37}$K)$_{\rm theory}$=7.5 efm$^2$.  The experimental value is larger in magnitude, but within 1$\sigma$ error of the shell-model result.  

$Q(^{37}$K) was remeasured using a newly-developed $\beta$-ray detecting nuclear quadrupole resonance ($\beta$ NQR) technique, to improve the precision of the previous measurement and to clarify the possible deviation from shell-model expectations.  The new result, $Q(^{37}$K) = 10.6 $\pm$ 0.4 efm$^2$, is well away from the shell-model result calculated in the $sd$ model space.  An extension of the shell-model space to include both $sd$ and $fp$ shells also does not account for the large $Q$ observed for the $^{37}$K ground state.  A possible variation of effective charges from the traditional values was explored to explain the discrepancy between theory and experiment.  
 
\section{Experiment}
\label{experiment}
The $Q(^{37}$K) measurement was performed at National Superconducting Cyclotron Laboratory (NSCL) at Michigan State University.  $^{37}$K ions were produced from a primary beam of $^{36}$Ar accelerated to 150 MeV/nucleon by the coupled cyclotrons and impinging on a 564 mg/cm$^2$ $^9$Be target.  A charge pick-up reaction ($^{36}$Ar + p $\rightarrow$ $^{37}$K) was employed to produce $^{37}$K because a large nuclear polarization is expected at the peak of the momentum-yield curve of the fragment \cite{groh03}.  The primary beam was steered at an angle of $+ 2^\circ$ relative to the normal beam axis at the production target in order to produce the polarized beam.  $^{37}$K ions were separated from other reaction products by the A1900 fragment separator \cite{morrissey97}.  The full angular acceptance ($\pm 2.5^\circ$) of the A1900 was used.  An achromatic wedge (1200 mg/cm$^2$ Al) was placed at the second intermediate dispersive image to separate $^{37}$K based on relative energy loss in the wedge.  The central momentum of the $^{37}$K beam was selected with a 1\% momentum acceptance to optimize the polarization.  A typical counting rate of $^{37}$K ions at the experimental apparatus was 50 particles/s/pnA and 40 pnA of primary beam was available.  The major contamination in the secondary beam following the A1900 was stable $^{36}$Ar, which did not adversely impact the $\beta$-NMR measurement.  

The polarized $^{37}$K ions were delivered to the $\beta$-NMR apparatus \cite{mantica97} and implanted into a tetragonal KH$_2$PO$_4$ (KDP) single crystal.  The crystal was 2 mm thick and tilted by 45$^{\circ}$ relative to the $\beta$-detector surface.  An external magnetic field of $H_0$ = 0.45 T was applied parallel to the direction of polarization and orthogonal to the beam direction to provide the Zeeman splitting field and to maintain the polarization in KDP.  $^{37}$K decays to the daughter $^{37}$Ar by emitting $\beta^+$ rays with a half life of 1.226 s.  The branching ratio to the ground state ($I^{\pi}$ = 3/2$^+$) is 98.2\%.  The maximum $\beta$-ray energy is 5.13 MeV.  The asymmetry parameter for direct $\beta$ decay to the $^{37}$Ar ground state is $A$ = $-$0.57.  $\beta$ rays from the stopped $^{37}$K ions were detected by a set of plastic scintillator telescopes placed at 0$^\circ$(u) and 180$^\circ$(d) relative to the external field direction.  The counting rate between u and d counters is asymmetric for a polarized source.  The angular distribution,
\begin{equation}
\label{eq:angulardistribution}
W(\varphi) \sim 1+AP\cos{\varphi},
\end{equation} 
depends on the polarization $P$, $A$ and the angle $\varphi$ between the direction of momentum of decaying $\beta$ ray and the polarization axis.  

The KDP crystal used for implantation had its c-axis set parallel to the external magnetic field.  The electric field gradient, $q$, at the K lattice site in KDP is axially symmetric (asymmetry parameter $\eta = 0$) and parallel to the c-axis.  The magnetic sub-level energies, $E_m$, of the $^{37}$K implanted in KDP are given in this condition by 
\begin{equation}
E_m = -g\mu_NH_0m + \frac{h\nu_Q}{12}(3\cos^2\theta-1)\{3m^2-I(I+1)\},
\label{eq:energy}
\end{equation}
where $m$ is the magnetic quantum number, $g$ is the nuclear $g$ factor, $\nu_Q = 3eqQ/\{2I(2I-1)h\}$ is the normalized electric coupling constant, where $eqQ/h$ is the electric coupling constant, and $\theta$ is the angle between the c-axis and external magnetic field. Eq. (\ref{eq:energy}) is given to the first-order of $eqQ/h$, taking the electric interaction as a perturbation to the main magnetic interaction.  Since the $^{37}$K nucleus implanted into KDP has a large electric interaction, the electric interaction can not be considered as a perturbation to the magnetic interaction.  Higher-order terms of the electric interaction have to be considered.  However, Eq. (\ref{eq:energy}) and (\ref{eq:frequency}) derived later may be used for the system with $\theta = 0$ and $\eta = 0$, where the first order perturbation calculation gives the exact transition frequencies.  The first term in Eq. (\ref{eq:energy}) indicates the $2I + 1$ magnetic sublevels separated by a fixed energy value due solely to the magnetic interaction (Zeeman splitting).  These sublevels are further shifted by the electric interaction and the energy spacing between adjacent sublevels is no longer constant.  The 2$I$ separate transition frequencies that appear due to the electric interaction are determined as
\begin{equation}
f_{m-1\leftrightarrow m} = \nu_L - \frac{\nu_Q}{4}(3\cos^2\theta-1)(2m-1).
\label{eq:frequency}
\end{equation}
The transition frequencies correspond to the energy difference between two adjacent energy levels in Eq. (\ref{eq:energy}), since the allowed transitions have $\Delta m = \pm1$.  Here $\nu_L = g\mu_NH_0/h$ is the Larmor frequency.   

The NMR signal is obtained from a double ratio defined as 
\begin{equation}
R = \left[\frac{W(0^\circ)}{W(180^\circ)}\right]_{\rm off}\Bigg/\left[\frac{W(0^\circ)}{W(180^\circ)}\right]_{\rm on}\sim 1+2A\Delta P,
\label{eq:udratio}
\end{equation} 
where the subscript, off(on), stands for without(with) RF applied and $\Delta P$ is the measured change in polarization between RF on and off conditions.  The NMR signal can be most effectively searched by applying the 2$I$ transition frequencies in Eq. (\ref{eq:frequency}) simultaneously, as a function of the electric coupling constant.  Such application ensures total destruction ($\Delta P=P_0$, where $P_0$ is the initial polarization produced in the nuclear reaction) of the polarization at the actual electric coupling constant and makes the measurement efficient.  The NMR signal for a nucleus with $I$ = 3/2 is 10 times bigger for $\Delta P = P_0$ than that obtained by applying a single transition frequency, where only a partial destruction of $P_0$ can be achieved.  A one day measurement with the multiple frequency technique would be at least a 100 days measurement with the single frequency technique.  In this calculation, a linear distribution of populations in $E_m$ is assumed as $a_m - a_{m-1} = \epsilon$ and $\sum^{I}_{m=-I}a_m = 1$, where $\epsilon$ is a constant.  This multiple frequency version of the $\beta$-NMR technique is known as the $\beta$-ray detecting nuclear quadrupole resonance ($\beta$ NQR) technique and is discussed in detail elsewhere \cite{minamiso07}.

\begin{figure}
\begin{center}
\includegraphics[width=12cm,keepaspectratio,clip]{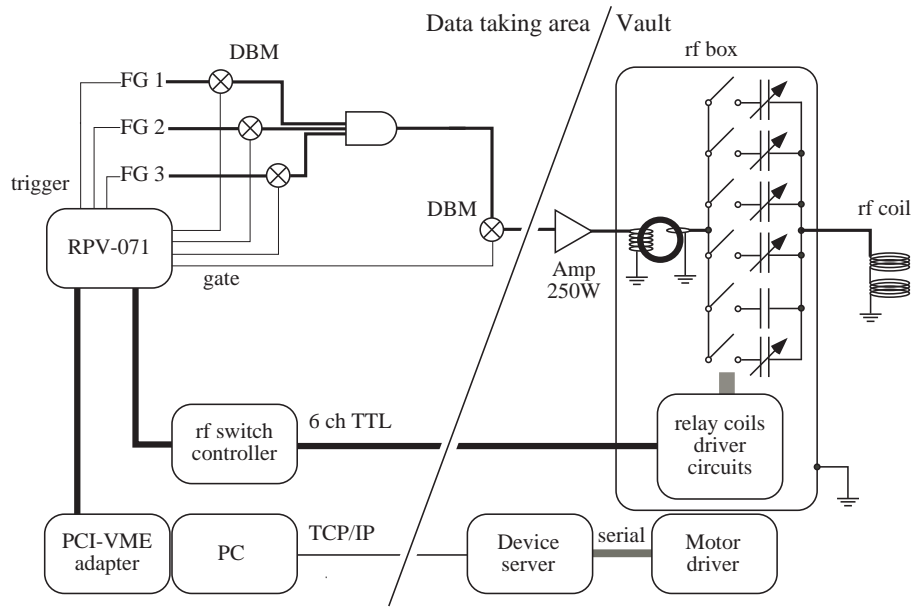}
\end{center}
\caption{Schematic illustration of the $\beta$-NQR system at NSCL.  One of the RF signals generated by the function generators is selected by a gate and sent to the amplifier.  The amplified signal is then applied to an RF coil through an LCR resonance circuit.  One of five variable capacitors and a fixed capacitor is selected by a fast relay switch to satisfy the resonance condition.  A pulse pattern generator, RPV-071, controls the function generators, the timing of DBM gating, the fast switching relay system and the pulsing of the short-lived radioactive beam.}
\label{fig:rfcircuit}
\end{figure}

A schematic of the $\beta$-NQR system at NSCL is shown in Fig. \ref{fig:rfcircuit}.  An RF signal generated by a function generator (labeled FG 1) is selected by a gate (labeled DBM) and sent to an RF amplifier.  The amplified signal is then applied to an RF coil, which is part of an LCR resonance circuit, with an impedance matching transformer and one of five variable capacitors and one fixed ($\sim$ 10000 pF) capacitor.  After 100 ms irradiation time, the frequency from FG 2 is selected by a gate signal and sent to the same LCR resonance circuit.  A different capacitor, which has been tuned to the second frequency to satisfy the LCR resonance condition, is selected by the fast switching relay system.  The system ensures sufficient power for any set of three transition frequencies over the expected search region of $eqQ/h = $ 1.9 $\sim$ 4.0 MHz, so that a high transition probability is achieved.  The transition frequency search range spans from 0.465 MHz to 2.465 MHz, as deduced from Eq. (\ref{eq:frequency}). 

The electric coupling constant of $^{37}$K in KDP was searched by applying three transition frequencies to the $^{37}$K ions.  The three transition frequencies for each electric coupling constant calculated using Eq. (\ref{eq:frequency}) were repeatedly applied, in sequence, through the RF coil.  Each signal was frequency modulated (FM = $\pm$ 20 kHz) to cover a certain region of the electric coupling constant.  The signal was frequency modulated at a rate of 50 Hz, and each signal was applied for 100 ms.  The oscillating magnetic field strength was $\sim$ 0.6 mT.  The $^{37}$K ions were continuously implanted into KDP and data were collected for consecutive RF on and off periods of 30 s each.

\section{Result}
\label{result}
\begin{figure}
\begin{center}
\includegraphics[width=12cm,keepaspectratio,clip]{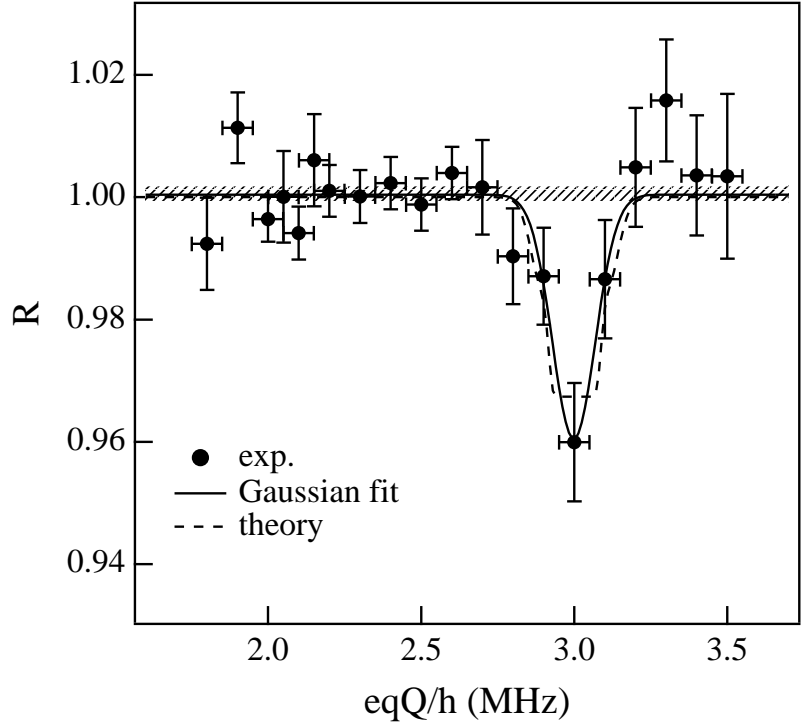}
\end{center}
\caption{Quadrupole resonance spectrum of $^{37}$K in KDP.  The NMR signal is plotted as a function of the electric coupling constant.  The solid circles are the data and the x-axis value of each point corresponds to a set of three transition frequencies.  The horizontal bar of each point is the range of electric coupling constant covered by the frequency modulation of the applied frequencies.  The shaded area is the base line of the resonance obtained from the data outside the resonance region.  The solid line is a Gaussian fit to the data and the dashed line is the theoretical line shape.}
\label{fig:37kinkdp}
\end{figure}

The $\beta$-NQR spectrum of $^{37}$K in KDP is shown in Fig. \ref{fig:37kinkdp}, where the NMR signal, $R$, is plotted as a function of the electric coupling constant.  The solid circles are the experimental data and the electric coupling constant of each point corresponds to a set of three transition frequencies, as noted earlier.  The horizontal bar on each point is the range of electric coupling constant covered by the frequency modulation of the applied frequencies.  The shaded area is the baseline of the resonance obtained from the data outside the resonance region.  The solid line is a Gaussian fit to the data.  The centroid of the fit gives the electric coupling constant of $^{37}$K in KDP:
\begin{equation}
\left|\frac{eqQ(^{37}{\rm K})}{h}\right| = 2.99 \pm 0.07({\rm stat.}) \pm 0.02({\rm syst.}) {\rm \ \ MHz}.
\label{eq:couplingconst}
\end{equation}
The dashed line is a theoretical line shape taking into account the frequency modulation of each transition frequency and the distribution of the electric field gradient $\Delta q/q$.  The latter broadens the natural line width and may lead to a different centroid.  The $\Delta q/q$ in KDP at the K lattice site is not known but  several percent is expected in ionic crystals \cite{minamisono84}.  It is difficult to fit  the theoretical line shape to data with $\Delta q/q$ as a fitting parameter since the data around the resonance are too sparse.  Therefore, the theoretical line shape was obtained with an assumption of $\Delta q/q$ = 3\%, giving reasonable reproduction of the data.  The systematic error in the coupling constant reflects the difference between the centroids obtained with the Gaussian and theoretical line shape fits.  

The electric coupling constant of $^{39}$K in KDP was measured by the conventional NMR technique as $|eqQ(^{39}{\rm K})/h|=1.69 \pm 0.01$ MHz \cite{seliger94}.  The ratio of the electric quadrupole coupling constants of $^{37}$K and $^{39}$K in KDP was determined to be:
\begin{equation}
\left[\frac{eqQ(^{37}{\rm K})}{h}\Bigg/\frac{eqQ(^{39}{\rm K})}{h}\right]_{\rm KDP} = \frac{Q(^{37}{\rm K})}{Q(^{39}{\rm K})} = 1.77 \pm 0.04.
\label{eq:eqqratio}
\end{equation}   
$Q(^{37}$K) can be extracted from Eq. (\ref{eq:eqqratio}) if $Q(^{39}$K) is known.  Ab initio calculations were performed to obtain the electric field gradient of the $3p^53d4s$ $^4F_{9/2}$ state in the $^{39}$K atom \cite{sundholm93}.  Applying the electric field gradient to the electric coupling constant of the $^4F_{9/2}$ state \cite{sprott68}, $Q(^{39}{\rm K}) = 6.01 \pm 0.15$ efm$^2$ was deduced.  This $Q(^{39}$K) was used to deduce $Q(^{37}$K) reported here.  The method to extract $Q$ from the measured $eqQ/h$ with a precisely calculated $q$ has been successfully applied to many atomic systems \cite{tokman98}.  Other reported values of $Q(^{39}$K) obtained from the 4$P_{3/2}$, 5$P_{3/2}$ or 6$P_{3/2}$ hyperfine coupling constants of the $^{39}$K atom \cite{arimondo77} with the Sternheimer correction \cite{sternheimer71}, and from molecular microwave spectra combined with calculations of electric field gradient \cite{kello98} are available as well.  The latter gives $Q$ values, which are consistent with and as precise as the ab initio calculations.  The values obtained with Sternheimer correction are subject to rather large error bars due to the accuracy of the correction \cite{arimondo77}.   However, these values of $Q$ are in good agreement within quoted errors and do not increase the systematic error of the present result.

$Q(^{37}{\rm K})$ was extracted from Eq. (\ref{eq:eqqratio}) and $Q(^{39}$K) discussed above as
\begin{equation}
Q(^{37}{\rm K}) = 10.6 \pm 0.4 {\rm\ \ efm}^2.
\label{eq:k37qmom}
\end{equation}
The error includes the statistical and systematic errors that resulted from the line shape of the $\beta$-NQR spectrum presented in Fig. \ref{fig:37kinkdp}.  A temperature dependence of the electric field gradient at the K lattice site in KDP was reported in Ref. \cite{seliger94}.  The temperature dependence was not considered for the $Q(^{37}$K) reported here, since its contribution is smaller than the uncertainty in the present result.   

The present value of $Q(^{37}$K) is consistent and more precise than the previous value, $Q(^{37}{\rm K})$ = 11 $\pm$ 4 efm$^2$, which was extracted from the hyperfine coupling constant of the $4p$ $^2P_{3/2}$ state of $^{37}$K atom, $B_{P_{3/2}}(^{37}{\rm K}) = 5.4 \pm 1.8$ MHz reported in Ref. \cite{behr97}, with the $4p$ $^2P_{3/2}$ coupling constant of $^{39}$K atom, $B_{P_{3/2}}(^{39}{\rm K}) = 2.83 \pm 0.13$ MHz \cite{arimondo77} and $Q(^{39}$K) = 6.01 $\pm$ 0.15 efm$^2$ evaluated with a calculated electric field gradient \cite{sundholm93}.

\section{Discussion}
\label{discussion}
\begin{figure}
\begin{center}
\includegraphics[width=12cm,keepaspectratio,clip]{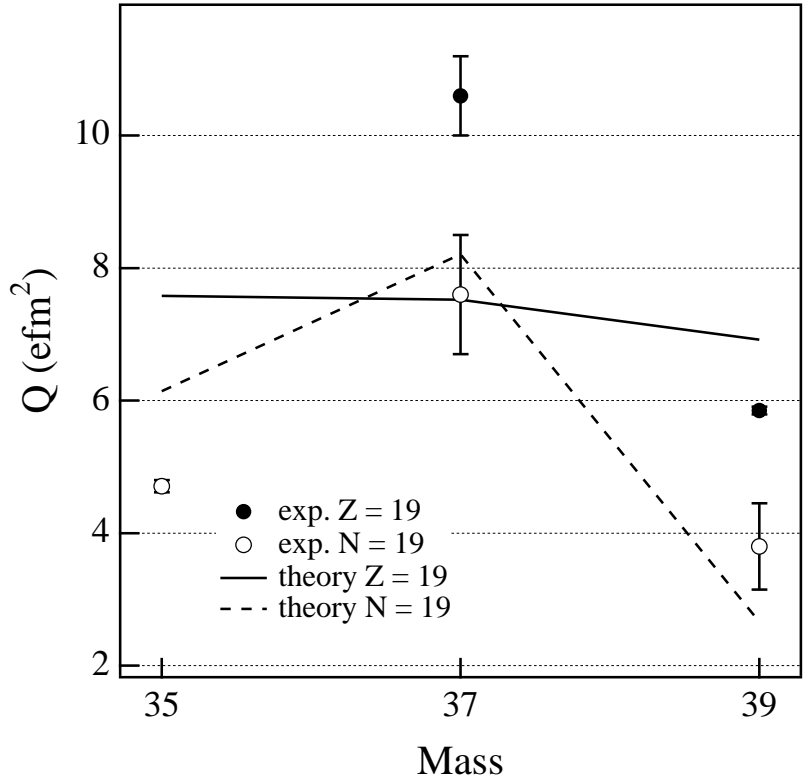}
\end{center}
\caption{Experimental and theoretical quadrupole moments of the K isotopes in the $sd$ shell and their mirror nuclei.  The solid circles and line are the data and theory of the K isotopes, respectively.  The open circles and dashed line are the data and theory of their mirror partners, respectively.}
\label{fig:qmomtheory}
\end{figure}

Theoretical calculations were performed using the OXBASH shell-model code \cite{brown} in the $sd$-model space with the USDA interaction \cite{brown06} and Woods-Saxson single-particle wave functions.  Theoretical quadrupole moments were calculated with $Q_{\rm theory} = e_pQ_0^p + e_nQ_0^n$, where $Q_0^{p(n)}$ is the bare quadrupole moment of the proton (neutron) and $e_p$ and $e_n$ are the effective charges for the proton and neutron, respectively.  Values of $e_p$ = 1.3$e$ and $e_n$ = 0.5$e$ were used for the shell-model results.  These effective charges are typical for $sd$ shell nuclei \cite{sagawa84, brown88}.  The experimental and theoretical results are summarized in Table 1 and shown in Fig. \ref{fig:qmomtheory} for the K isotopes ($Z$ = 19) and their mirror nuclei (the $N$ = 19 isotones).  Calculations with the USD \cite{wildenthal83} or the USDB \cite{brown06} interactions gave results within the expected theoretical uncertainty and therefore only the results of the calculations performed with USDA interaction are discussed here.  The quadrupole moments of the mirror nuclei were obtained by exchanging the bare quadrupole moments of the proton and neutron, assuming charge symmetry.  The theoretical result $Q(^{37}$K)$_{\rm theory}$ = 7.5 efm$^2$ is significantly smaller than the present experimental value [Eq. (\ref{eq:k37qmom})].  The theoretical ratio [$Q(^{37}$K)/$Q(^{39}$K)]$_{\rm theory}$ = 1.1 also disagrees with the experimental result [Eq.(\ref{eq:eqqratio})] as well.  Finally, the experimental trend of K isotopes is poorly reproduced by the shell model results as depicted in Fig. \ref{fig:qmomtheory}.

The smaller theoretical ratio of $Q(^{37}$K) to $Q(^{39}$K) than deduced from experiment is understood as follows.  An increase in $Q(^{37}$K)$_{\rm theory}$ is expected when two neutrons are removed from $^{39}$K.  This is realized from the $Q_0^n$ value reported in Table 1.  However, a decrease in $Q_0^p$ is also calculated between $^{39}$K and $^{37}$K, which is nearly half the increase in $Q_0^n$.  The reduction of $Q_0^p$ is attributed to two neutrons coupled to 2$^{+}$ in the $d_{3/2}$ shell.  The wave function of the ground state of $^{37}$K is approximated by 
\begin{align}
\label{eq:37kwavefunction}
\left|^{37}{\rm K}(3/2^+)\right> &\sim \alpha \left|\left[\pi d^{-1}_{3/2}\times\nu(^{\rm 38}{\rm Ca}: 0^+)\right]^{3/2}\right>\notag\\ 
&+ \beta \left|\left[\pi d^{-1}_{3/2}\times\nu(^{\rm 38}{\rm Ca}: 2^+)\right]^{3/2}\right>,
\end{align}  
where $^{38}{\rm Ca}:0^+$ and $^{38}{\rm Ca}:2^+$ represent the ground and first excited states, respectively.  Eq. (\ref{eq:37kwavefunction}) can be rewritten as $\left|^{37}{\rm K}(3/2^+)\right> \sim \alpha \left|1\right> + \beta \left|2\right>$.  The amplitudes of these states, $\alpha^2$ and $\beta^2$, are calculated to be 0.76 and 0.16, respectively.  The $\left|^{37}{\rm K}(3/2^+)\right>$ covers $\sim$92\% of all the possible wave functions in the $sd$ model space.  The contribution from each term to the $Q_0^p$ can be evaluated as $\alpha^2\left<1||\widehat{Q}_p||1\right> = 1\times\alpha^2\left<d_{3/2}^{-1}||\widehat{Q}_p||d_{3/2}^{-1}\right>$ and $\beta^2\left<2||\widehat{Q}_p||2\right> = -0.6\times\beta^2\left<d_{3/2}^{-1}||\widehat{Q}_p||d_{3/2}^{-1}\right>$, where $\widehat{Q}_p$ is the quadrupole operator.  The factors 1 and $-$0.6 in the righthand side of each equation are the phase factors given by the initial and final state spins.  The opposite sign of the $\left<2||\widehat{Q}_p||2\right>$ term to that of the $\left<1||\widehat{Q}_p||1\right>$ term reduces the $Q^p_0$ of $^{37}$K compared to that of $^{39}$K, where the $\left<1||\widehat{Q}_p||1\right>$ term is dominant.  The neutron configuration in $^{37}$K counteracts the $Q_0^p$ and the reduction results in the small increase of the overall $Q(^{37}$K)$_{\rm theory}$ relative to $Q(^{39}$K)$_{\rm theory}$.  Consequently, the experimental ratio, namely the trend in $Q$, would be difficult to reproduce in this framework of the shell model, even if other interactions are considered.

Cross shell (2p-2h) excitations were considered for protons and neutrons with the Monte Carlo Shell Model (MCSM) in the $sd$ and $fp$ model space with the SDPF-M interaction \cite{utsuno99}.  The fractions of valence protons and neutrons excited to the $fp$ shell in the calculation are 0.29 and 0.28 for $^{37}$K and 0.32 and 0.42 for $^{39}$K, respectively.  The resulting [$Q(^{37}$K)/$Q(^{39}$K)]$_{\rm theory}$ = 1.1 and $Q$($^{37}$K)$_{\rm theory}$ = 7.47 efm$^2$ are still significantly smaller than experimental values.  The MCSM results are surprisingly similar to the OXBASH result ( [$Q(^{37}$K)/$Q(^{39}$K)]$_{\rm theory}$ = 1.1 and $Q$($^{37}$K)$_{\rm theory}$ = 7.52 efm$^2$, respectively), which were limited to the $sd$ model space.  Although $\sim$ 30\% of all the nucleons in the MCSM calculation are excited to the $fp$ shell, the $Q(^{37}$K)$_{\rm theory}$ is not increased, compared to that calculated in the $sd$ model space.  The contribution of the 2p-2h excitations to $Q$($^{37}$K)$_{\rm theory}$ can therefore be considered negligible. 

The experimental result may be better reproduced by the shell-model calculations discussed above with an adjustment of the effective charges.  A quadrupole deformation of the valence orbitals will lead to a self-consistent polarization of the core.  The deformation is described as the coupling of 2$\hbar \omega$, $I^\pi$=2$^+$ excitations of the core to the valence orbits \cite{sagawa84} and is represented within the model space as effective charges for protons and neutrons.  As discussed in the introduction, there is evidence of reduced effective neutron charges in the B isotopes near the neutron dripline \cite{ogawa04}.  But no precedent has been established for changes in effective charges in neutron-deficient nuclei like $^{37}$K.

The experimental ratio between $Q(^{37}$K) and $Q(^{39}$K) may be used as a guide in estimating effective charges.  The ratio can be represented as:
\begin{equation}
\label{eq:theoreticalratio}
\frac{Q(^{37}{\rm K})}{Q(^{39}{\rm K})} = \frac{Q_0^p(^{37}{\rm K})}{Q_0^p(^{39}{\rm K})} + \frac{e_n}{e_p}\frac{Q_0^n(^{37}{\rm K})}{Q_0^p(^{39}{\rm K})},
\end{equation}
where the $N$ = 20 neutron core in $^{39}$K couples to 0$^+$.  The ratio of the effective charges can be extracted as $e_n/e_p \sim$ 1.14, using Eq. (\ref{eq:eqqratio}) and the bare quadrupole moments reported in Table 1.  The typical effective charges $e_p = 1.3e$, $e_n = 0.5e$ give $e_n/e_p \sim 0.4$.  An increase in $e_n$ and/or a decrease in $e_p$ is required to reproduce the experimental $Q(^{37}$K)/$Q(^{39}$K) [Eq. (\ref{eq:eqqratio})].

The suggestion of alternate effective charges is also seen in the experimental quadrupole moments of mirror nuclei near $^{37}$K.  $^{39}$K and $^{39}$Ca have one proton hole and one neutron hole, respectively, in doubly-magic $^{40}$Ca.  The simple shell-model configurations of these nuclei give the effective charge as the ratio of experimental and bare quadrupole moments for protons ($^{39}$K) and neutrons ($^{39}$Ca).  The effective charges obtained in this approach are $e_p$ $\sim$ 1.13$e$ and $e_n$ $\sim$ 0.71$e$.  The increase in $e_n$ and decrease in $e_p$, relative to the typical values, lead to [$Q(^{37}$K)/$Q(^{39}$K)]$_{\rm theory} \sim 1.3$ and $Q(^{37}$K)$_{\rm theory}$ = 7.9 efm$^2$.  These values are increased relative to the shell-model results with the typical values of effective charges, but still remain smaller than the experimental values.  The $Q$ evaluated with these new effective charges derived from the $A$ = 39, $T$ = 1/2 mirror partners are summarized in Table 1.  

The driving force for the variation of effective charges may lie in the interaction between the valence neutrons and core protons.  The effective charges can be expanded to include the isoscalar, $e_{\rm pol}^{(0)}$, and isovector, $e_{\rm pol}^{(1)}$, polarization charges arising from the virtual excitation of the isoscalar and isovector giant quadrupole resonances of the core: 
\begin{equation}
e_p = 1e+e_{\rm pol}^{(0)}-e_{\rm pol}^{(1)}, \ \ \ e_n = e_{\rm pol}^{(0)}+e_{\rm pol}^{(1)}.
\label{eq:polcharge} 
\end{equation}
The $e_{\rm pol}^{(0)}$ is well determined from the $E$2 matrix elements between low-lying states \cite{brown88}, which is dominated by the isoscalar component, and can be derived from typical effective charges as $e_{\rm pol}^{(0)} \sim 0.4e$.  Therefore, using $e_n/e_p \sim 1.14$ obtained by considering the ratio of experimental $Q$ of $^{37}$K and $^{39}$K given in Eq. (\ref{eq:theoreticalratio}), a value $e_{\rm pol}^{(1)} \sim 0.6e$ is determined, much larger than the value $e_{\rm pol}^{(1)} \sim 0.1e$ from the typical effective charges, $e_p = 1.3e$ and $e_n = 0.5e$.  The large $e_{\rm pol}^{(1)}$ indicates stronger coupling to the isovector quadrupole giant resonance.  It is also noted here that such a large $e_{\rm pol}^{(1)}$ has also been discussed for the $A$ = 51 mirror nuclei \cite{rietz04}, which reside in the $fp$ shell.  

An alternative way to evaluate $Q$ is by a collective model approach.  The ground state wave function of $^{37}$K given in Eq. (\ref{eq:37kwavefunction}) may also be considered as a single-hole configuration in the $d_{3/2}$ shell interacting with the 2$^+$ vibrational state of the $^{38}$Ca core.  The influence of particle-core coupling can be calculated as a correction to the $Q$ by taking into account the effect induced by the collective $E$2 operator.  A simple relation between $Q$ in the coupled particle-core system and the $Q$ of the single particle configuration can be derived \cite{neyens97} as:
\begin{align}
\label{eq:particlecore}
Q(^{37}{\rm K};3/2^+) & = Q(\left|\pi d^{-1}_{3/2}\right>; 3/2^+)e_p^{\rm eff}\\
& = Q(\left|\pi d^{-1}_{3/2}\right>; 3/2^+)\left\{1 + \frac{2}{r^2}\frac{4\pi}{15}\left<r\frac{dV}{dr}\right>\frac{B(E2; 0^+_1 \rightarrow 2^+_1)}{ZeR_0^2\hbar\omega_2}\right\}.\notag
\end{align}  
The neutron-dependent proton effective charge, $e_{p}^{\rm eff}$, can be derived from the experimental $B(E2)$ and excitation energy $\hbar\omega_2 = E_{2_1^+}$ of the first-excited $2^+_1$ state in the $^{38}$Ca core.  Eq. (\ref{eq:particlecore}) can be evaluated using the values, $Q(\left|\pi d^{-1}_{3/2}\right>; 3/2^+) \sim 5.2$ fm$^2$, $\left<r^2\right>_{d_{3/2}}$ $\sim$ 13 fm$^2$, $\left<rdV/dr\right> \sim$ 40 MeV, $E_{2^+_1}$ = 2.21 $\pm$ 0.01 MeV and $B(E2; 0_1^+ \rightarrow 2_1^+) = 96 \pm 21 e^2$fm$^4$ \cite{cottle99}, to obtain $Q(^{37}{\rm K}) = 8.7 \pm 0.8$ efm$^2$.  This collective approach using empirical data gives a larger value of $Q$ than the shell model.  The agreement with the present experimental result is improved, however, the error is large.  A $B(E2)$ with higher precision would be useful for further discussion.  

The large experimental $Q(^{37}$K) and relatively small proton separation energy of $^{37}$K, $S_p = 1.858$ MeV, would be an indication of extended valence proton orbit or even the development of a proton halo \cite{minamisono92, sumikama06}.  In such a proton halo system, $e_p$ should converge to 1$e$ due to a smaller probability of the valence protons to polarize the core.  Though the experimental $Q(^{37}$K)/$Q(^{39}$K) may indicate a smaller $e_p$, a corresponding increase in $e_n$ is essential to better reproduce $Q(^{37}$K) and the ratio $Q(^{37}$K)/$Q(^{39}$K) as discussed above.  Also, the mean-square charge radius of $^{37}$K relative to that of $^{39}$K is known \cite{behr97} and does not show evidence of an extended distribution of protons.  Therefore, the large $Q(^{37}$K), the small $e_p$ and the small proton separation energy of $^{37}$K are not attributed to an extended proton orbit or a proton halo.  The angular momentum of the valence nucleon ($d_{3/2}$) and the large atomic number ($Z$ = 19) are expected to suppress an extended proton orbit in $^{37}$K, due to large centrifugal and Coulomb barriers, respectively.  

\section{Summary}
\label{summary}
A newly-developed $\beta$-NQR system was used at NSCL to measure the charge distribution of the neutron-deficient nucleus $^{37}$K.  The electric quadrupole coupling constant of $^{37}$K, implanted in a tetragonal KDP single crystal, was measured to be $|eqQ/h|$ = 2.99 $\pm$ 0.07 MHz.  Together with the known electric quadrupole coupling constant of $^{39}$K in KDP and quadrupole moment of $^{39}$K, the quadrupole moment of $^{37}$K was deduced as $|Q(^{37}{\rm K})|$ = 10.6 $\pm$ 0.4 efm$^2$.  The present result is consistent with and significantly more precise than the previous value (11 $\pm$ 4 efm$^2$), evaluated from the hyperfine coupling constant measured by laser spectroscopy.  Shell-model results in the $sd$ or the $sd$ and $fp$ model spaces are not able to reproduce the large $Q(^{37}$K) value.  A increased neutron effective charge and a decreased proton effective charge relative to the typical values in the $sd$ shell are required for a better agreement between experiment and theory.  Such changes in the effective charges for the $Q$ of neutron-deficient $^{37}$K indicate that the well bound valence neutrons strongly polarize the core protons.  More substantial coupling to the isovector giant resonance is inferred, beyond that indicated by the typical effective charges for the $sd$ shell nuclei.  

\section*{Acknowledgement}
This work was supported in part by the National Science Foundation, Grant PHY06-06007.  The authors would like to express their thanks to the NSCL operations staff for providing the primary and secondary beams and the electronics group  for their help in developing the RF system.



\newpage

\begin{table}
\begin{center}
\caption{Experimental and theoretical quadrupole moments of the $Z$ = 19 isotopes and $N$ = 19 isotones.  The $Q_0$ values, calculated with shell model using the USDA interaction, are bare quadrupole moments without effective charges.  Typical effective charges, $e_p$ = 1.3$e$ and $e_n$ = 0.5$e$ in the $sd$ shell and effective charges obtained from $Q_{\rm exp}$ for the $A$ = 39 mirror pair, $e_p$ = 1.13$e$ and $e_n$ = 0.71$e$, were used to calculate $Q_{\rm theory}$.} 

\begin{tabular}{lcccccr}
\hline 
nucleus & $Q^p_0$ & $Q^n_0$ & \multicolumn{2}{c}{$Q_{\rm theory}$ (efm$^2$)} & $Q_{\rm exp.}$ (efm$^2$) & Ref.\\[-0.2in]
 & & & $e_p$, $e_n$ & $e_p$, $e_n$ & \\[-0.2in]
 & & & 1.3$e$, 0.5$e$ & 1.13$e$, 0.71$e$ & \\
\hline 
$^{39}$K & 5.32 & 0 & 6.92 & 6.01 & 6.01 $\pm$ 0.15 & \cite{sundholm93}\\
$^{37}$K & 3.94 & 4.80 & 7.52 & 7.86 & 10.6 $\pm$ 0.4 & present \\
$^{35}$K & 4.71 & 2.91 & 7.58 & 7.39 & -  & \\
$^{39}$Ca & 0 & 5.32 & 2.66 & 3.78 & 3.8 $\pm$ 0.65 & \cite{stone05} \\
$^{37}$Ar & 4.80 & 3.94 & 8.21 & 8.22 & 7.6 $\pm$ 0.9 & \cite{stone05}\\
$^{35}$S & 2.91 & 4.71 & 6.14 & 6.63 & 4.71 $\pm$ 0.09 &\cite{stone05} \\
\hline
\end{tabular}
\end{center}
\end{table}

\newpage

\newpage

\newpage

\end{document}